# Towards a Layered Architectural View for Security Analysis in SCADA Systems

Zhendong Ma, Paul Smith, Florian Skopik
Safety & Security Department
Austrian Institute of Technology
2444 Seibersdorf, Austria
Email: {zhendong.ma, paul.smith, florian.skopik}@ait.ac.at

*Abstract*—Supervisory Control and Data Acquisition (SCADA) systems support and control the operation of many critical infrastructures that our society depend on, such as power grids. Since SCADA systems become a target for cyber attacks and the potential impact of a successful attack could lead to disastrous consequences in the physical world, ensuring the security of these systems is of vital importance. A fundamental prerequisite to securing a SCADA system is a clear understanding and a consistent view of its architecture. However, because of the complexity and scale of SCADA systems, this is challenging to acquire. In this paper, we propose a layered architectural view for SCADA systems, which aims at building a common ground among stakeholders and supporting the implementation of security analysis. In order to manage the complexity and scale, we define four interrelated architectural layers, and uses the concept of viewpoints to focus on a subset of the system. We indicate the applicability of our approach in the context of SCADA system security analysis.

## I. INTRODUCTION

Supervisory Control and Data Acquisition (SCADA) systems are computer systems that monitor and control industrial facilities and processes. Many critical infrastructures such as power grid, traffic management, gas and water facilities are supported and controlled by SCADA systems. In SCADA systems, computer applications as well as human operators in the control center collect measurements from remotely connected sensors and send commands to actuators in the field according to predefined process models. Hence, SCADA systems help to extend the activities in the cyber space to the physical world. Because of the critical nature of the physical devices and services they monitor and control, SCADA systems are a ready target for cyber attacks. Historically, most SCADA systems had been separated from other networks and used proprietary protocols, hardware and software. However, in recent years, due to technological development and economic considerations, SCADA systems are becoming increasingly interconnected and tend to use Commercial-Off-The-Shelf (COTS) IT products as well as open standards. For example, in many cases, SCADA systems are connected to an organization's enterprise networks, which are in-turn connected to the Internet. Consequently, SCADA systems have to face the same vulnerabilities and threats that plague normal IT systems. In addition, connectivity and the use of COTS products makes it easier for an attacker to understand and search for weakness in the system. Stuxnet [1] is a well-exposed wakeup call on the imminent danger facing SCADA systems and our critical infrastructures.

Identifying the vulnerabilities and threats and protecting SCADA systems against cyber attacks is of vital importance. However, due to their characteristics this is not straightforward. SCADA systems tend to be complex, for example, because of the heterogeneity of the systems and communication technologies they use, and large-scale in nature. Besides, SCADA systems have very strict real-time requirements, and the lifecycle of SCADA systems tends to be much longer than normal IT systems. This makes changing and hardening the infrastructure for security purposes difficult. A summary of the challenges of securing SCADA networks is discussed by Igure *et al.* in [2], and a set of best practice solutions are given by [3]. Despite this being an acknowledged problem, securing SCADA systems continues to be a significant challenge [4], [5].

Clearly, to secure SCADA systems a systematic approach must be taken. Security processes that makes use of existing best practices and guidelines are a promising approach because they are more likely to be adopted by today's critical infrastructure asset owners. A fundamental building block in any of the security process is *security analysis*, which aims at identifying assets, vulnerabilities, and the associated threats and potential attacks. A prerequisite of security analysis is to gain a systematic and comprehensive understanding of the SCADA system under consideration. Due to their scale, complexity, and heterogeneity, a consistent view rarely exist (cf. Section II). To this end, we propose a novel way to organize and describe SCADA systems and their environments and contexts using architectural layers. As often required in software design-level risk analysis, the layered architectural view is envisioned to assist security analysis by slicing and organizing SCADA system to different technological concerns and layers of abstraction in order to *build up a consistent "forest-level" view of the target system at a reasonably high level* [6].

Our architectural view is comprised of four interrelated layers: *asset*, *communication*, *service*, and *organization* layer,

which capture important aspects of a SCADA system from a security perspective. The established layered architectural view can be applied during the process of vulnerability and threat analysis, risk assessment and secure architecture design. To manage the scale and complexity of SCADA systems, we also make use of *viewpoints* to make intersections through the layers when they are applied to specific security-related tasks.

The rest of this paper is organized as follows: in Section II, we describe related work and motivate the need for a novel architectural view for SCADA systems. Section III describes our approach to establish the architectural view, including a set of principles associated with the four layers. We also introduce the use of viewpoints to manage SCADA system complexity and scale. To indicate the applicability, in Section IV we discuss its usage in the context of a set of security-related processes. Section V concludes the paper and points out our further work.

## II. Related work

A consistent architectural view establishes a framework for understanding the target system, the entities within the system and their relationships, and system environments. When undertaking security processes that involve different organizations and personnel, a reference architectural view helps to maintain consistency and common consensus among the participants.

Most common approaches so far are to model a SCADA system in accordance with its network topology. For example, a typical SCADA architecture for power grid includes field devices connected to the SCADA network, which is connected to a corporate network [7]. This architectural approach is adapted in [8] for developing simulation tools for SCADA systems security. Such a topological view of the architecture is also used in [3], [2], [9]. In our approach, details of the network infrastructure are captured in a distinct layer that describes how data may flow in a SCADA system; additionally, we would like to capture other aspects that are equally important from a security perspective.

Other approaches focus on the software services within SCADA systems. The VIKING reference architecture [10] models services and data flow, and their relationship to the network topology. An architecture meta-model includes three components: data-flow, service, and zone. In our approach, these aspects are modeled across different layers, making it more straightforward to identify and analyze the security aspects of each of the entities. Coupled with viewpoints, our layers can be collapse to consider multiple layers in security analysis. The ability to identify and classify interdependencies within SCADA systems is important for security. Berg and Stamp [11] propose a system reference architecture by applying *Object-Role Modeling* [12] in order to model data, functionality and internal interdependencies of SCADA systems. In their approach, an object represents the features and properties of a system entity, and a role annotates the relationship between the objects. The objects in a SCADA system are grouped into four levels – infrastructure equipment, SCADA field equipment, systems and plant control centers, and automation oversight. This organization is based on a mixture of function and network topology. As mentioned earlier, we separate infrastructure and networking aspects into distinct layers in order to make their analysis more approachable. Using viewpoints, we can also collapse the layers in case of carrying out the object role modeling proposed by Berg and Stamp.

A slightly different approach is to first identify SCADA devices and then logically group their functions into abstraction layers. The ISA S99 standard [13] proposes to create a reference architecture from the entities identified as assets within an organization, and build the architecture model according to the specifics of the organization. ISA S99 maps the functional components of a SCADA system into five architectural levels: physical process, local or basic control, supervisory control, operations management, and enterprise systems. Our approach follows the same principle, i.e., we propose to first build an asset layer as the basis for other layers.

The authors of [14], [15] propose a SCADA reference architecture with added architectural components for security and resilience. In their approach, the entire SCADA architecture is modeled as a wide area network connecting several local area networks. SCADA entities are distributed within the boundaries of each of the local networks. In essence, such an architectural view focuses only on the hostile environment connecting the local networks.

Based on our study of existing work, it can be observed that there is no consensus on how to model SCADA system architectures, including what should be modeled. However, system architectures tend to be modeled to reflect network segmentation. Indeed, there is no single solution that provides us with a comprehensive and tailorable view of a SCADA system architecture for use when applying security analysis.

## III. Layered Architectural View

A reference architecture should capture the essence of the architecture of a collection of systems [16]. Our architectural view is structured into four *layers*, which can be considered in the context of arbitrary *viewpoints*. This arrangement is summarized in Figure 1. Each layer is intended to group system components and aspects for security analysis. A layer in the architecture consists of entities that are typically considered discrete – for example, the communication layer includes aspects from layer two and three of the OSI reference model, and the asset layer describes physical and logical entities. Since each SCADA system is unique, e.g., using a range of components and subsystems from different vendors configured in different ways, it is impossible to have an architecture model that captures all peculiarities of different SCADA systems working in different organizations and domains. Therefore, our proposal here is meant to be an architectural template, based on which specific SCADA system architectural views can be derived and instantiated.

To further manage the complexity and scale of SCADA systems, we make use of the concept of *viewpoints*. A viewpoint is *a technique for abstraction using a selected set of*

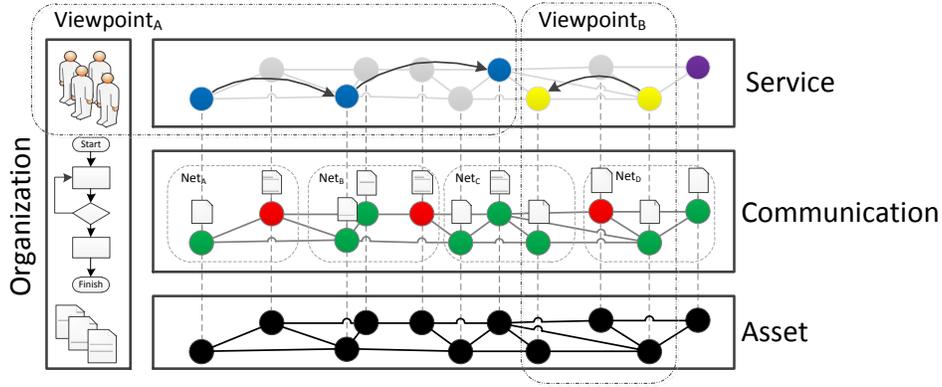

Fig. 1. Layered SCADA system architectural view

*architectural concepts and structuring rules, in order to focus on particular concerns within that system* [17]. We define a viewpoint to be an arbitrary view of a SCADA system that focuses on a subset of the system. A viewpoint can include system components from the same architectural layer as well as those from different layers. A viewpoint may be defined, for example, in order to understand the vulnerability associated with a new aspect of a SCADA system or to determine the implementation of a high-level security policy via processes, software and hardware.

*A. Asset Layer*

The asset layer includes entities such as hardware, software, and data of a SCADA system that is usually considered as the IT asset of an organization. Hardware of a SCADA system are physical devices as well as the communication links that connect them. Devices in a SCADA system can include those typically associated with enterprise networks, such as workstations and servers, and those that are related to SCADA systems, such as field devices, including Intelligent Electronic Devices (IDEs), Remote Terminal Unit (RTUs), Programmable Logic Controller (PLCs) and Distributed Control Systems (DCS). The field devices monitor meter readings and equipment status and control end devices such as sensors and actuators. Usually, the wired and wireless communication links connect the devices into the following topologies: geographically distributed field devices are connected over various communication links (e.g., dial-up telephone, leased line, power line, radio, and Wide Area Network (WAN)) to control centers in SCADA networks; the SCADA network is connected to a company's corporate network, and the corporate network is further connected to the Internet; firewalls are used to separate and protect different networks. Software include operating systems, databases, and application software. Data is generated and processed by hardware and software components in SCADA systems. In the asset layer, the software and data are associated with specific hardware. Figure 2 illustrates an example of asset layer.

Components in the asset layer should be relatively straightforward to identify. For example, a critical infrastructure asset owner typically has detailed information on each of the hardware devices and how they are connected – e.g., maintained in an asset management system – as well as the software installed on that hardware and the data exchanged at the I/O ports or APIs. The components can be specified using common IT asset specification methods like the "Specification for asset identification" from NIST [18], which defines a data model with asset types such as software, database, network, and service etc.

*B. Communication Layer*

The way data can be transmitted and the means of realizing these data flows, e.g., using various protocols and services, is important for SCADA system security – the communication layer aims to describe this. This understanding can be applied when carrying out a threat analysis to determine the reachability of critical assets from remote networks, both internal and external. Furthermore, understanding which protocols are being used can identify vulnerabilities in the SCADA system. Building on top of the entities described in the asset layer, the characteristics of three main classes of entity are described in this layer: (1) communication *enablers*, (2) communication *inhibitors*, and (3) communication *end points*, as shown in Figure 3.

Communication enablers include devices such as network hubs, switches and routers, for example; a further form of enabler includes the means of interconnecting these devices, physically or virtually, e.g., with the use of Virtual Local Area Networks (VLANs). In short, communication enablers describe and implement how data should flow through the SCADA system. In contrast to the various communication enablers, inhibitors curb the flow of data through the SCADA system, and typically take the form of so-called middleboxes, such as firewalls and Network Address Translation (NAT) devices. An intrinsic communication inhibitor in SCADA systems can come from the heterogeneity of the protocols that are used. For example, TCP/IP is the ubiquitous protocol arrangement in enterprise networks; whereas in control networks a wide-range of protocols are used, such as DNP3 and Modbus. The communication end points are entities sending or receiving data.

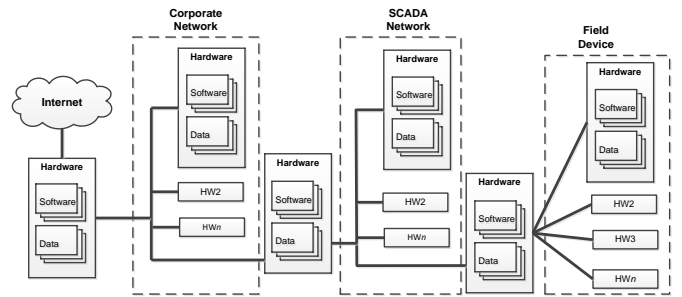

Fig. 2. Asset layer

With respect to the communication enablers and inhibitors three main attributes should be enumerated: (1) the addressing layout (or configuration) of the SCADA system, e.g., in terms of network and subnetwork address ranges; (2) the protocols and services that are used; and (3) the state associated with these protocols and services. Collectively, these three items define how data can flow through the SCADA system. Enabling protocols include those used for routing (e.g., OSPF and BGP) and their state includes items such as their configuration and routes held in routing tables. Inhibiting services include those running on firewalls and their associated rule set.

Furthermore, there are typically a number of services that support communication in a SCADA system, such as management and measurement services (e.g., SNMP and Netflow data collection, respectively). Determining the existence and configuration of these services is important, for example, to understand vulnerabilities at this layer.

## C. Service Layer

The service layer models the software services, applications, or functions in SCADA systems and the data exchanged among the services. We use the term "service" in a broad sense to denote software components that encapsulate and provide certain functionality. Consequently, databases, authentication servers, Web servers and application servers are considered services. A service can be a composition of standalone services which provides customized functionalities and business applications. A service can be implemented and deployed using numerous techniques, ranging from low-level embedded systems components to application software and flexible service mashups and orchestration engines. The data flows models the data exchanges among the services.

An example service layer is illustrated in Figure 4, which is a simplified version adapted from [10]. Services like sensor and actuator send measurement data and receive process commands to and from SCADA server through front end. HMI services are the interfaces of the operators to the SCADA server. Historian stores the historical data. A web browser in the corporate network can access these data. The SCADA application server provides various control center applications such as power projects and overview monitoring. A Geographic Information System (GIS) service provides GIS data to data engineering server, which defines data structures and views for various services in the SCADA network. Communication server allows remote client to have terminal access to SCADA server for faster and more efficient maintenance work and information acquiring.

Adapting some of the Service Oriented Architecture (SOA) terminology, we can describe a service with the following attributes: (1) *Service Descriptions* define the capabilities and functional properties of a service, as well as the communication endpoint and operations supported by the service; (2) *Interactions and Data Contracts*, which define the schema used for exchanged messages as well as the protocol of interaction; (3) *Fault Handling Procedures* provide information in case of failures and undeliverable services; (4) *Service Level Agreements* hold important information about guaranteed quality criteria, such as availability, accuracy, responsiveness and so on. While these definitions are more feasible for business software in cooperate networks, the same concepts are also applicable to services in SCADA networks. For instance, in order to use a deployed firmware-controlled sensor, one needs to know the protocol and applicable messages to interact with this sensor; furthermore, what operations this sensor provides (type of measurement, value ranges, operating modes), how faults are signaled, and what level of service in terms of availability or accuracy this sensor guarantees.

Describing software as services with encapsulated functions makes it easier to model SCADA systems with legacy and proprietary software components. With a certain level of abstraction, the service layer enable us to focus on security of distributed information systems. For example, on a service layer, it is possible to identify vulnerabilities of *one* particular service, e.g., because of using a weakly implemented version, or applying an inappropriate configuration which might cause open backdoors. Furthermore, it is also possible to identify vulnerabilities due to compositional aspects. For example, a standalone secure service B being composed with a vulnerable service A can be exploited, because A provides an attack surface to B, e.g., a backdoor in A is used to reach B. On the other side, an application can also be vulnerable resulting from bad design, even if the single services are secure. This happens, for example, if services are not composed in the right manner, such as a data provisioning service reachable from the Internet lacks a composition with a proper authentication service. However, in order to ensure the usage of appropriate compositions and security configuration of single software components, we also need to model corporate security policies on an organizational level.

## D. Organization Layer

The organization layer consists of relevant people and their activities, as described by organizational processes and policies. As an orthogonal layer, the entities in the organization layer are related in different ways to the other layers. This includes end-users of services provisioned by SCADA systems as well as maintenance personnel keeping the whole infrastructure up and running. People on an organization layer have predefined privileges (e.g., roles and access rights) on peer objects, including hardware, network configurations, services and data entities. A majority of an organization's activities can be described by business processes. Business processes can potentially span numerous departments across organization boundaries. Business requirements and corporate rules affect the execution of business processes, i.e., the order and context of tasks being performed. Furthermore, in most processes coordination through and intervention of humans is required, which makes people not only system end-users, but integral parts of the whole system architecture. When executing such processes, predefined tasks are performed by different stakeholders in series. This requires the delegation of privileges among people, for instance, the ownership of data objects depending on the current task. Finally, the execution of processes is influenced by security policies and guidelines, especially, how close they are lived and applied in the business context. Thorough monitoring of user actions and review with respect to these security-relevant artifacts can reveal weaknesses in corporate procedures.

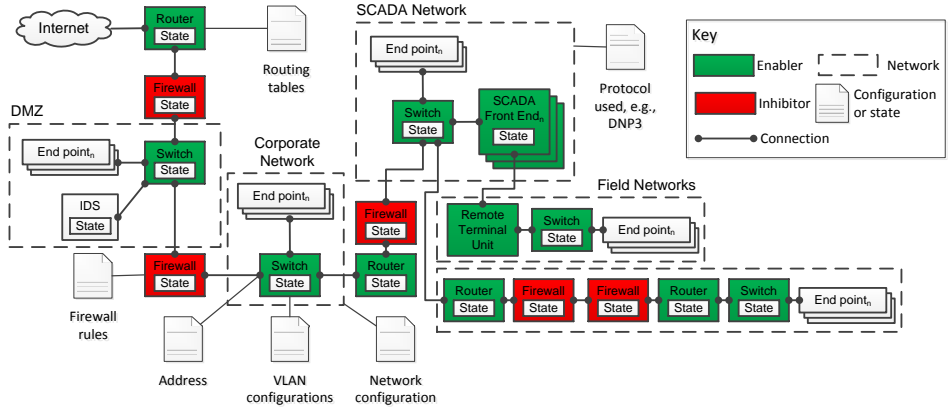

Fig. 3. Communication layer

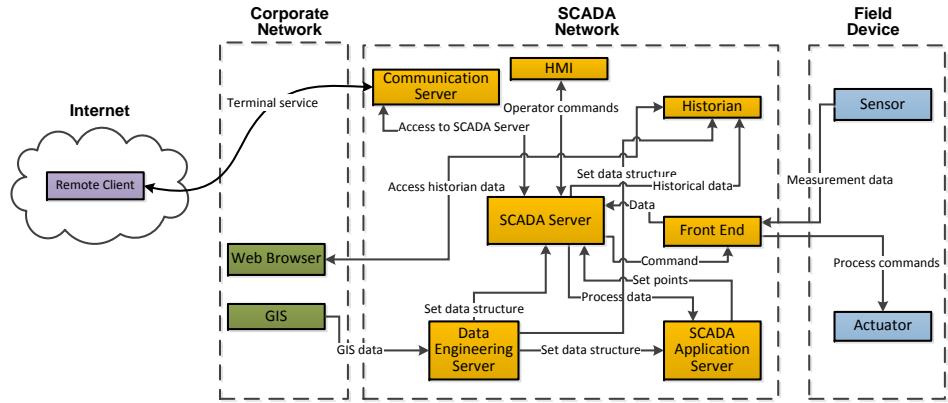

Fig. 4. Service layer

We regard security policies, which describe security administration rules and enforcement hierarchy, as an integral part of the organization layer. Security policies include those for general IT systems, such as information security and risk management policies, as well as specific policies for SCADA systems such as platform security, communication security, and application security policies [19].

*E. Viewpoints*

As mentioned earlier, viewpoints are intended to provide a focused view of a subset of the architectural layers. A viewpoint can be horizontally aligned to a layer, or vertically intersect different layers.

Viewpoints can be arbitrarily defined based on the security process that is being carried out. For example, a security team may wish to validate how an organizational level security policy is implemented from a technology and processes perspective. To do this they may define a set of viewpoints from the perspective of the different security policies under examination, which cuts across the organizational, service and communication layers, for example. This could reveal how a security policy is implemented in service access control mechanisms and firewall policies.

A key challenge when defining viewpoints is determining their scope, i.e., identifying the SCADA system entities that are relevant with respect to a proposed viewpoint. In our refer-ence architecture the interdependencies and other relationships between entities in the various layers will be modeled, e.g., the dependency of a service and its composition on the communication and physical infrastructure. We are exploring how a viewpoint might be semi-automatically abstracted from these relationships as a graph, for example, as a *polytree* – a form of directed acyclic graph that has a single path between two vertices – whose edges are determined based on conditional probabilities, such as that of being in a compromised state.

## IV. Security Analysis

The precise nature of the instantiation of the architectural view, e.g., which entities will be enumerated and their attributes in accordance with the architectural template, will depend on its application to the security analysis of specific SCADA systems. A security team can make use of abstract representations of the architectural view to manage the complexity and scale of the system, and to introduce some automation to the process. We foresee a number of applications of our reference architecture. For example, we can use the framework from Schaeffer *et al.* [20] for developing and evaluating so-called resilience patterns that describe the configuration of various mechanisms, e.g., firewalls and anomaly detection systems, that can be used to detect and mitigate well-known attacks, such as DDoS attacks. Their framework makes use of simulations to evaluate candidate resilience patterns; our reference architecture could be used to support the realisation of simulation models in this context. In a more formal way, we can apply the layered architectural view for attack modeling in smart grid proposed by Chen *et al.* [21] that makes use of Petri-nets. In their approach, "low-level" Petri-nets are created by domain experts that describe attacks in detail for sub-domains of a smart grid, e.g., attacks on smart meters. Then the low-level attack descriptions are merged with "high-level" Petri-nets that abstract details of an attack, and focus on important places, i.e., attack states. Common places in the two types of Petri-net are merged by identifying the transitions and places described by a common model description language. Using the layered architectural view, we can support this attack modeling approach – for example, the systematic identification of low-level Petri-nets could be done on a per-layer basis, and viewpoints could be defined that identify places and their attributes across multiple layers. Furthermore, the lexicon of the model description language could be derived from the attributes contained in an instantiation of our reference architecture.

## V. Conclusion and Future Work

Since SCADA systems are the IT backbone of many critical infrastructures, security analysis that identifies vulnerabilities, threats and attacks is an important task for securing and protecting critical infrastructures against cyber attacks. Establishing a consistent architectural view of the target system shoiuld be the first step in any security analysis. In this paper, we proposed a layered architectural view to support the implementation of security analysis. As a novel way to organize and describe architectural information and to manage complexity and scale, we model SCADA systems in four layers: asset, communication, service, and organization layer. In addition, we introduced the concept of architectural viewpoints, which enables us to have a focused view on a subset of the system of interest during security analysis. We are aware that it is a challenging task to enforce a unanimous view on SCADA system architecture among various stakeholders. The proposed architectural view is an attempt to establish and maintain a consistent view on the system architecture during security processes. With abstract and focused presentations in the architectural view, we envision that more theoretical and formal methods, as well as automation techniques can be developed and applied for security analysis in SCADA systems.

There are several directions for our future work: we will apply the architectural view in a research project that aims at preventing and protecting critical infrastructure against cyber attacks to gain more practical experiences of the feasibility of our approach. The architectural view captures interdependency among components within a SCADA system, which can be exploited for identifying interrelated vulnerabilities and threats. SCADA systems are complex and typically in large scale. Tool support for efficient instantiation of the architectural view will be another objective in our future work.